  \documentclass[final,5p,times,twocolumn]{elsarticle}

\usepackage{amssymb}
\usepackage{mathrsfs}

\journal{Physics Letters B}

\begin{document}

\begin{frontmatter}



\title{Residual interaction 
in Second RPA with density-dependent forces. 
Rearrangement terms}

\author{M. Grasso}

\address{Institut de Physique Nucl\'eaire,
 Universit\'e Paris-Sud, IN2P3-CNRS, F-91406 Orsay Cedex, France}

\author{D. Gambacurta}
\address{Dipartimento di Fisica e Astronomia and INFN, Via Santa
  Sofia 64, I-95123 Catania, Italy}

\author{F. Catara}
\address{Dipartimento di Fisica e Astronomia and INFN, Via Santa
  Sofia 64, I-95123 Catania, Italy}

\begin{abstract}
We derive the expression for the residual interaction to be used 
in the framework of the Second RPA with  
density-dependent forces. The adopted procedure 
is based on a variational approach. 
It is found that the residual interaction to be used 
in RPA and beyond RPA matrix elements is not the same as far as the 
rearrangement terms are concerned. 
A detailed derivation of the matrix elements coupling 1 particle-1 hole 
with 2 particle-2 hole and 2 particle-2 hole among themselves has been 
done and the corresponding rearrangement terms are shown here. 
This formal result indicates that both the currently used prescriptions, namely (i) using the same 
type of rearrangement terms in RPA and beyond RPA matrix elements or (ii) 
neglecting the rearrangement terms in beyond RPA matrix elements, are not 
correct. 
\end{abstract}


\end{frontmatter}






The RPA is a widely used  microscopic approach adopted to study collective 
modes which can be interpreted in terms of vibrations. In nuclear physics, 
the employment of this model with effective density-dependent interactions 
 allows to successfully describe giant resonances and low-lying excitation 
modes. The second RPA (SRPA) is a natural extension of the RPA where 2 
particle-2 hole (2$p$-2$h$) excitations are included together with the 
usual RPA 1 particle-1 hole ($1p-1h$) configurations, providing in this way 
a richer description of the excitation modes. On the other hand, the 
manifestation  of new exotic modes  in unstable nuclei has produced an 
increasing interest in the context of 
nuclear structure and the inclusion of complex
configurations coupled among themselves may play a crucial role in the 
description
of these phenomena. 
Moreover, some low-energy excitations like 
the first 0$^+$ and 2$^+$ states in magic nuclei are not well reproduced 
by the standard RPA model (they are predicted too high in energy) 
because 
beyond 1$p$-1$h$ configurations are needed to describe them.
Finally, SRPA would allow to study in a proper way the double 
phonon excitation modes that are experimentally well known in nuclei.  
In all these cases, interesting information may be provided by the application 
of  SRPA.

The SRPA equations  are 
well known since several decades  (see, e.g., \cite{yann}). 
However, up to very recently, they 
have never been fully and self-consistently solved due to 
the heavy numerical effort that they require. 
Some approximations have been 
adopted in the past, 
namely the SRPA equations have been reduced to a simpler 
second Tamm-Dancoff model (see for instance \cite{ho,kn,ni1,ni2}) 
and/or the equations have been solved with uncorrelated 
2$p$-2$h$ states, i.e. the residual interaction terms in the matrix that couples $2p2h$  configurations among themselves have been 
neglected (diagonal approximation)
 \cite{ad,sc1,sc2,dr1,dr2,yann2,yann3}. 
Very recently, 
the SRPA equations have been solved for some closed-shell nuclei using 
an interaction derived from the Argonne V18 potential (with the Unitary 
Correlation Operator Method) 
\cite{roth}, for small metallic clusters in the jellium approximation 
\cite{gamba} and for the magic nucleus $^{16}$O with the Skyrme interaction 
\cite{gamba1}.

The most currently employed 
phenomenological interactions in mean field 
approaches are density-dependent forces of Skyrme or Gogny type. 
It is well known that, with density-dependent forces, the residual 
interaction used to evaluate the RPA matrices $A$ and $B$ contains  
rearrangement terms generated by the derivative with respect to the density 
of the mean field hamiltonian (second derivative of the energy density 
functional). Moreover, rearrangement terms are present also in  
the single particle energies appearing in the diagonal part of the 
matrix $A$. Due to the density dependence of the 
interaction, the mean field is changed and the single particle energies 
are thus also modified by the variations of 
the density.  

 When dealing with the SRPA problem with density-dependent 
forces, a first formal aspect to consider is the determination of the 
residual interaction which has to be used in beyond RPA matrix elements. 
To our knowledge, this aspect has not yet been clarified in the 
literature. 

In Ref. \cite{waro} the issue of the rearrangement terms has been
studied  in the context of shell model calculations while in Ref. \cite{ad1}
 a prescription is introduced for the rearrangement  
terms appearing in matrix elements  beyond the standard RPA ones. However,
 in more recent calculations \cite{ad2}, 
the same authors have not actually used that prescription and have not 
included 
those rearrangement terms. 

This work differs from that of Ref. \cite{ad1}
for the following reasons. The ground state of the system is described as 
a superposition of 1$p$-1$h$ and 2$p$-2$h$ configurations build on top of the 
Hartree-Fock (HF) state, (see Eq. (\ref{eq2})). The coefficients 
describing these configurations are used as variational parameters and the 
SRPA equations are obtained, as shown in Ref. \cite{prov}, by minimizing the 
expectation value of the Hamiltonian with respect to them. On the
 contrary, in Ref. \cite{ad1} the 2$p$-2$h$ degrees of freedom are not taken 
into account explicitly.  
 In the case of density dependent forces, the RPA equations can be derived  
in the small amplitude limit of the Time Dependend Hartre-Fock (TDHF) method 
which allows to define the residual interaction as the second derivative of 
the energy density functional with respect to the density. This implies that
 the rearrangement terms appear and they are related to the first and second 
derivative of the interaction with respect to the density. Therefore, 
differently from Ref. \cite{ad1},  we consider an expansion of 
the interaction up to the second order in the density. This allows 
obtaining the correct definition of the rearrangement terms in the RPA 
matrices. Finally, the analysis done in \cite{ad1} deals only 
with the rearrangement terms in the matrix elements between 
the 1$p$-1$h$ and 2$p$-2$h$ configurations while the coupling of the 
2$p$-2$h$ configurations among themselves is not studied.

In this work, we discuss the issue of the residual interaction in SRPA 
using the variational derivation of the SRPA equations proposed by 
Providencia \cite{prov}, extending this derivation to the case of a density dependent
force. 
As done in \cite{prov}, we express the ground state $|\Psi >$ as
\begin{eqnarray}
|\Psi > = exp S |\Phi >~,
\label{eq1}
\end{eqnarray}
where $|\Phi>$ is the HF ground state and $S$ is a linear superposition of 
1$p$-1$h$ and 2$p$-2$h$ configurations
\begin{eqnarray}
S = \sum_{ph} C_{ph} a^{\dag}_p a_h + \frac {1} {2}  
\sum_{pp'hh'} C_{pp'hh'} a^{\dag}_p a^{\dag}_{p'}  a_{h'} a_h~.
\label{eq2}
\end{eqnarray}
considering thus a different and more general wave function than the one used 
in \cite{ad1}.

Let us denote with $h$, $i$, $j$, $k$ and $l$ hole states and with 
$m$, $n$, $p$ and $q$ particle states.  We introduce: 
\begin{eqnarray}
\hat{C}_{\alpha \beta \gamma \delta} = C_{\alpha \beta \gamma \delta} 
-  C_{\alpha \beta \delta \gamma}~.
\label{eq3}
\end{eqnarray}

We consider the case of a density-dependent interaction  and the Hamiltonian is thus written  in the following form:
\begin{eqnarray}
H=\sum_{\alpha \beta} t_{\alpha \beta} a^{\dag}_{\alpha} a_{\beta} + 
\frac {1} {4} \sum_{\alpha \beta \gamma \delta}
\hat{V}_{\alpha \beta \gamma \delta} (\rho)  a^{\dag}_{\alpha} 
a^{\dag}_{\beta}  
a_{\delta} a_{\gamma}~,
\label{H}
\end{eqnarray}
where:
\begin{eqnarray}
\hat{V}_{\alpha \beta \gamma \delta} (\rho) = 
V_{\alpha \beta \gamma \delta} (\rho)   - 
V_{\alpha \beta \delta \gamma} (\rho)~.
\label{eq5}
\end{eqnarray}

The one body density $\rho$ matrix can be expanded around the HF density $\rho^{(0)}$ 

\begin{eqnarray}
 \rho_{\alpha \beta} =  
<\Psi| a^{\dag}_{\beta} a_{\alpha} |\Psi> = <\Phi| 
e^{S^{\dag}} a^{\dag}_{\beta} a_{\alpha} e^S |\Phi>
 \nonumber \\
 = <\Phi| (1+S^{\dag} + \frac {1} {2} S^{\dag 2}+... ) 
a^{\dag}_{\beta} a_{\alpha}  (1+ S + \frac {1} {2} S^2 +...) | \Phi>  
\nonumber \\
 \sim \rho_{\alpha \beta}^{(0)} + 
<\Phi| a^{\dag}_{\beta} a_{\alpha} S + S^{\dag} 
a^{\dag}_{\beta} a_{\alpha} |\Phi> + \nonumber \\
<\Phi| \frac {1} {2} a^{\dag}_{\beta} a_{\alpha} S^2 + 
S^{\dag} a^{\dag}_{\beta} a_{\alpha} S + 
\frac{1} {2} S^{\dag 2} a^{\dag}_{\beta} a_{\alpha}|\Phi> =
\rho_{\alpha \beta}^{(0)} + \delta \rho_{\alpha \beta} 
\label{eq7}
\end{eqnarray}
where we have truncated at the quadratic terms. 
Hence, the variation of the density is given by the sum of a 
linear $\delta \rho^{(1)}$ and a quadratic 
$\delta \rho^{(2)}$ contributions written as:
\begin{eqnarray}
\delta \rho^{(1)}_{\alpha \beta} = 
<\Phi| a^{\dag}_{\beta} a_{\alpha} S + S^{\dag} 
a^{\dag}_{\beta} a_{\alpha} |\Phi> 
\label{eq8}
\end{eqnarray}
and 
\begin{eqnarray}
\delta \rho^{(2)}_{\alpha \beta} = 
<\Phi| \frac {1} {2} a^{\dag}_{\beta} a_{\alpha} S^2 + S^{\dag} a^{\dag}_{\beta} a_{\alpha} S + 
\frac{1} {2} S^{\dag 2} a^{\dag}_{\beta} a_{\alpha}|\Phi>~.
\label{eq8_1}
\end{eqnarray}

In particular, one can find that:
\begin{eqnarray}
\delta \rho^{(1)}_{hh'}= \delta \rho^{(1)}_{pp'} =0~;
\label{eq9}
\delta \rho^{(1)}_{ph}= C_{ph}~;
\label{eq10}
\delta \rho^{(1)}_{hp} = C^*_{ph}~;
\label{eq11}
\end{eqnarray}

\begin{eqnarray}
\delta \rho^{(2)}_{ph}= \sum_{mi}C^*_{mi} \hat{C}_{pmhi}~;
\label{eq12}
\delta \rho^{(2)}_{hp} = \sum_{mi} C_{mi} \hat{C}^*_{pmhi}~;
\label{eq13}
\end{eqnarray}
 
\begin{eqnarray}
\delta \rho^{(2)}_{hh'} = -\sum_{m} C^*_{mh} C_{mh'} - \frac {1} {2} 
\sum_{mni} \hat{C}^*_{mnih} \hat{C}_{mnih'}~;
\label{eq14}
\end{eqnarray} 

\begin{eqnarray}
\delta \rho^{(2)}_{pp'} = \sum_{i} C^*_{p'i} C_{pi} + \frac {1} {2} 
\sum_{mij} \hat{C}^*_{p'mij} \hat{C}_{pmij}~.
\label{eq15}
\end{eqnarray} 

The mean value of the Hamiltoninan in $|\Psi>$, $<H>$, can be written as in Eq. (8) of Ref. \cite{prov}: 
\begin{eqnarray}
<H>= <\Phi|H|\Phi> + \sum_{mi} (C^{*}_{mi} \lambda_{mi} (\rho) + C_{mi} \lambda_{im} (\rho)) +  \nonumber \\
\sum_{i<j,m<n} (\hat{C}^{*}_{mnij} \hat{V}_{mnij} (\rho) + \hat{C}_{mnij} \hat{V}_{ijmn} 
(\rho))  + F^{(2)} ~,
\label{meanH}
\end{eqnarray} 
where $F^{(2)}$ is the sum of several quadratic contributions in $C$ and $C^*$ (see Eq. (8) of Ref. \cite{prov} 
for details). 

In the above equation:
\begin{eqnarray} 
\lambda_{ab}(\rho) = t_{ab} + \sum_k \hat{V}_{akbk}(\rho)~.
\label{eq17}
\end{eqnarray}
which, in cases where the interaction is not density-dependent, 
define the single particle energies.
The $A$ matrices appearing in the SRPA equations are defined by
\begin{eqnarray}
A_{mi,pk } = \left[\frac {\delta^2 <H> } {\delta C^*_{mi} \delta C_{pk}}\right]_{0}\equiv A_{11};
\label{A11}
\end{eqnarray}

\begin{eqnarray}
A_{mi,pqkl } = \left[\frac {\delta^2 <H> } {\delta C^*_{mi} \delta \hat{C}_{pqkl}}\right]_{0} \equiv A_{12};
\label{A12}
\end{eqnarray}

\begin{eqnarray}
A_{mnij,pqkl } = \left[\frac {\delta^2 <H> } {\delta \hat{C}^*_{mnij} \delta \hat{C}_{pqkl}}\right]_{0} \equiv A_{22};
\label{A22}
\end{eqnarray}
which are evaluated at $C=C^{*}=\hat{C}=\hat{C}^{*}=0$. 
The corresponding $B$ matrices are obtained by substituting in the 
previous equations the derivative with respect to $C$ and $\hat{C}$ with the
derivative with respect to
$C^{*}$ and $\hat{C}^{*}$, respectively.
For example, the RPA $B$ matrix is
\begin{eqnarray}
 B_{mi,pk } = \left[\frac {\delta^2 <H> } {\delta C^*_{mi} \delta C^*_{pk}}\right]_{0}.
\label{B11}
\end{eqnarray}
When the interaction is not density-dependent, 
only the  $F^{(2)}$ term of Eq. (\ref{meanH}) gives some contributions, 
since the others are not quadratic in the $C$'s coefficients. 
The standard SRPA matrices are thus obtained 
(see Eqs (9)-(12) of Ref. \cite{prov}). If, on the contrary, the interaction 
is density-dependent, rearrangement terms appear and they are given by the 
first three terms of Eq. (\ref{meanH}). In the following we focus our 
attention only on the rearrangement terms that have to be included in 
SRPA matrices. 

As mentioned above, in the case of density dependent forces, RPA equations can 
be obtained in the small amplitude limit of TDHF and the single-particle 
energies contain only the first derivative of the interaction  with respect to 
the density, while both the first and the second derivatives  appear in the 
residual interaction. We thus expand $\hat{V}$ around the HF density 
$\rho^{(0)}$ and keep up to quadratic terms :
\begin{eqnarray}
&& \hat{V}_{\alpha \beta \gamma \delta} (\rho) \sim 
\hat{V}_{\alpha \beta \gamma \delta} (\rho^{(0)}) + 
\sum_{a b} \left[ \frac {\delta \hat{V}_{\alpha \beta \gamma \delta}} 
{\delta \rho_{ab}}\right]_{\rho=\rho^{(0)}} \delta \rho_{ab} \nonumber \\
&& + \frac {1} {2} 
\sum_{a b c d } \left[ \frac {\delta^2 \hat{V}_{\alpha \beta \gamma \delta} } 
{\delta \rho_{ab} \delta\rho_{cd}}\right]_{\rho=\rho^{(0)}} \delta \rho_{ab} \delta \rho_{cd} ~;
\label{eq6}
\end{eqnarray}
where 
\begin{eqnarray}
 \delta \rho_{\alpha \beta}=\delta \rho_{\alpha \beta}^{(1)}+\delta \rho_{\alpha \beta}^{(2)}.
\end{eqnarray}

It can be shown that, by using the above definition for the RPA matrices and 
the previous expansion for the interaction,  the usual RPA 
rearrangement terms are found.

For example, for the  $A_{11}$ matrix defined in Eq (\ref{A11}) we find
\begin{equation}
A_{mi,pk } = \delta_{ki}\epsilon_{mp}-\delta_{mp}
\epsilon_{ki}+\mathscr{V}_{mkip},
\label{A11-RPA}
\end{equation}

with the single particle energies given by

\begin{equation}
 \epsilon_{ab}  =\lambda_{ab}(\rho^{(0)})+\frac{1}{2}\sum_{kk'} 
\left[\frac {\delta \hat{V}_{k k' k k'}} 
{\delta \rho_{ba}} \right]_{\rho=\rho^{(0)}} \rho_{ba},  
\label{spenergy}
\end{equation} 
where $\rho_{ba}=\psi_b^* \psi_a$ and $\psi$ are the HF single particle wave 
functions. 
The rearrangement contribution in Eq. (\ref{spenergy}) comes out from the 
derivative of the first term appearing in the right side of Eq. (\ref{meanH}); 
in the derivation of Eq. (\ref{spenergy}) we have used that 
$\rho_{hh'}^{(0)}=\delta_{hh'}$.
In Eq. (\ref{A11-RPA}), the residual interaction  is 
\begin{eqnarray}
 \mathscr{V}_{mkip}=\hat{V}_{mkip} (\rho^{(0)}) + 
 \sum_{h} \left[ \frac {\delta \hat{V}_{m h i h}} 
 {\delta \rho_{kp}}\right]_{\rho=\rho^{(0)}} \rho_{kp} +\nonumber \\
 \sum_{h } \left[ \frac {\delta \hat{V}_{h k h p}} 
 {\delta \rho_{mi}}\right]_{\rho=\rho^{(0)}} \rho_{mi} 
  + \frac {1} {2} 
 \sum_{h h' } \left[ \frac {\delta^2 \hat{V}_{h h' h h'} } 
 {\delta \rho_{mi} \delta\rho_{kp}}\right]_{\rho=\rho^{(0)}} 
\rho_{mi} \rho_{kp},
\label{vres-a11}
\end{eqnarray}
where the first  term is the original interaction while all the others are the 
rearrangement terms. The first two come out from the $\lambda$ quantities 
appearing in Eq. (\ref{meanH}), while the last one derives 
from the mean value of the Hamiltonian in the HF state, i.e. the first term 
of Eq. (\ref{meanH}).   

Let us now focus our attention on the rearrangement terms appearing in the 
residual interaction of the matrix elements beyond the usual RPA.
First of all, it comes out that no rearrangement terms appear in the matrix 
$A_{22}$, Eq. (\ref{A22}), except for the contributions
\begin{eqnarray}
 A_{mnij,pqkl}^{(rearr)}=
+\frac{1}{2}\delta_{nq}\delta_{ik}\delta_{jl}\sum_{kk'} 
\left[\frac {\delta \hat{V}_{k k' k k'}} 
 {\delta \rho_{pm}}\right]_{\rho=\rho^{(0)}} \rho_{pm}   \nonumber \\
 +\frac{1}{2}\delta_{mp}\delta_{ik}\delta_{jl}\sum_{kk'} 
 \left[\frac {\delta \hat{V}_{k k' k k'}} 
 {\delta \rho_{qn}}\right]_{\rho=\rho^{(0)}} \rho_{qn} \nonumber \\
  -\frac{1}{2}\delta_{mp}\delta_{nq}\delta_{jl}\sum_{kk'} 
 \left[\frac {\delta \hat{V}_{k k' k k'}} 
 {\delta \rho_{ki}}\right]_{\rho=\rho^{(0)}} \rho_{ki} \nonumber \\
 -\frac{1}{2}\delta_{mp}\delta_{nq}\delta_{ik}\sum_{kk'} 
 \left[\frac {\delta \hat{V}_{k k' k k'}} 
 {\delta \rho_{lj}}\right]_{\rho=\rho^{(0)}} \rho_{lj}
\label{a22-r}
\end{eqnarray}
 generated by the first term of Eq. (\ref{meanH}). The above terms,  
together with the 
$\lambda$ quantities, give the correct single particle energies in the case 
of a density-dependent interaction, being thus completely consistent with the 
single particle energies appearing in the RPA matrix (\ref{A11-RPA}). 
No rearrangement terms  appear  in the $B_{22}$ matrix.

Let us consider now the matrices coupling 1$p$-1$h$ with 2$p$-2$h$ configurations: the same kind of contributions  discussed just above is found, that together with the first term appearing in Eq. 10 of ref. \cite{prov} gives a vanishing term due to the HF condition

\begin{equation}
 \epsilon_{kp}=\lambda_{kp}(\rho^{(0)})+\frac{1}{2}\sum_{kk'} 
\left[\frac {\delta \hat{V}_{k k' k k'}} 
 {\delta \rho_{pk}}\right]_{\rho=\rho^{(0)}} \rho_{pk} =0 .
\label{spea12}
\end{equation}

In addition to that, we have only one type of rearrangement terms in the 
residual interaction, given by: 

\begin{eqnarray}
A_{mi,pqkl }^{(rearr)} = \left[ \frac {\delta \hat{V}_{klpq}} 
 {\delta \rho_{im}}\right]_{\rho=\rho^{(0)}}\rho_{im }.
\label{A12-r}
\end{eqnarray}

A similar term is found in the $B_{12}$, namely 
\begin{eqnarray}
B_{mi,pqkl }^{(rearr)}=\left[ \frac {\delta \hat{V}_{klpq}} 
 {\delta \rho_{mi}}\right]_{\rho=\rho^{(0)}}\rho_{mi }.
\label{B12-r}
\end{eqnarray}

We recall that, in the case of a two-body non density-dependent interaction, 
the 
$B_{22}$ and $B_{12}$ are found to be zero. On the contrary, in the present 
case $B_{12}$ 
has a non vanishing contribution coming from the rearrangement terms.
This  can  be easily understood since the density-dependent 
part of the interaction 
mimics a three-body force and takes thus into account more complicated 
correlations.

Some conclusions can be thus drawn for the residual interaction of the 
matrix elements beyond RPA. 
A first kind of rearrangement terms appears in the matrix $A_{22}$  giving the 
right definition of the single particle energies, consistently with RPA. 
The same kind of terms appears also in the  $A_{12}$ matrix which 
finally gives a vanishing contribution due to the HF condition 
Eq. (\ref{spea12}).  
No  rearrangement terms appear instead in the residual interaction of the 
$A_{22}$ and $B_{22}$ matrices. 
The $B_{22}$ matrix is thus still equal to zero. 
A comment can be added concerning this last result, that, as a matter of fact, introduces an asymmetry between the $A_{11}$ and $A_{22}$ matrices, as far the residual interaction is concerned. This can be traced back to the fact that the interaction depends  only on the one-body density. A different situation would occur in the case of more general forces, for example depending also on the two-body density matrix.

In the matrices $A_{12}$ and $B_{12}$, rearrangement terms of the residual 
interaction are found and they are given by Eqs. (\ref{A12-r}) and 
(\ref{B12-r}), respectively. Their expression is  different from the one of 
the RPA matrices. Therefore, the same definition of the residual interaction 
as in standard RPA cannot be adopted in the matrix elements describing the 
coupling between
1 particle-1 hole with 2 particle-2 hole and 2 particle-2 hole among 
themselves. At the same time, completely neglecting the rearrangement terms 
in the matrix elements beyond RPA is not correct. 
A quantitative study of the terms (\ref{A12-r}) and (\ref{B12-r}) could be 
useful and it will be done in a future work.


\vspace{2cm}

We are grateful to J.-F. Berger, N. Pillet, P. Schuck and N. Van Giai  
for fruitful  and useful discussions.

\end{document}